# Identification of a Nematic Pair Density Wave State in $Bi_2Sr_2CaCu_2O_{8+x}$

Weijiong Chen[1,§], Wangping Ren[1,§], Niall Kennedy[1,2], M. H. Hamidian[3], S. Uchida[4], H. Eisaki[4], Peter. D. Johnson[1,5], Shane O'Mahony[2] and J.C. Séamus Davis[1,2,3,6]

1. *Clarendon Laboratory, University of Oxford, Oxford, OX1 3PU, UK.*
2. *Department of Physics, University College Cork, Cork T12 R5C, Ireland*
3. *Department of Physics, Cornell University, Ithaca, NY 14850, USA*
4. *Inst. of Advanced Industrial Science and Tech., Tsukuba, Ibaraki 305-8568, Japan.*
5. *CMPMS Department, Brookhaven National Laboratory, Upton, NY 11973, USA*
6. *Max-Planck Institute for Chemical Physics of Solids, D-01187 Dresden, Germany*
§   *These authors contributed equally to this project.*

ABSTRACT    Electron-pair density wave (PDW) states are now an intense focus of research in the field of cuprate correlated superconductivity. PDW's exhibit periodically modulating superconductive electron pairing which can be visualized directly using scanned Josephson tunneling microscopy (SJTM). Although from theory, intertwining the *d*-wave superconducting (DSC) and PDW order parameters allows a plethora of global electron-pair orders to appear, which one actually occurs in the various cuprates is unknown. Here we use SJTM to visualize the interplay of PDW and DSC states in $Bi_2Sr_2CaCu_2O_{8+x}$ at a carrier density where the charge density wave (CDW) modulations are virtually nonexistent. Simultaneous visualization of their amplitudes reveals that the intertwined PDW and DSC are mutually attractive states. Then, by separately imaging the electron-pair density modulations of the two orthogonal PDWs, we discover a robust nematic PDW state. Its spatial arrangement entails Ising domains of opposite nematicity, each consisting primarily of unidirectional and lattice commensurate electron-pair density modulations. Further, we demonstrate by direct imaging that the scattering resonances identifying Zn impurity atom sites occur predominantly within boundaries between these domains. This implies that the nematic PDW state is pinned by Zn atoms, as was recently proposed (Lozano et al, PHYSICAL REVIEW B 103, L020502 (2021)). Taken in combination, these data indicate that the PDW in $Bi_2Sr_2CaCu_2O_{8+x}$ is a vestigial nematic pair density wave state (J. Wardh and M. Granath arXiv:2203.08250v1).

The existence and phenomenology of PDW states are fundamental issues within the challenge to understand cuprate correlated superconductivity [1,2,3,4]. Their definitive characteristic is a periodically modulating electron-pair density which can now be visualized directly at the atomic-scale, by using scanned Josephson tunneling microscopy[5,6,7,8,9] (SJTM). This new capability allows one to explore both the microscopic electronic structure of the cuprate PDW state, and the interactions between it and the other electronic orders. Understanding the latter is essential, because intertwining the *d*-wave superconducting (DSC) and PDW order parameters allows a plethora of distinct electron-pair orders to appear[1,2,3]. But which one actually occurs in the various cuprates is unknown.

The simplest quantum condensates of electron pairs are defined by an homogeneous superconductive order-parameter

$$\Delta_0(\boldsymbol{r}) = \Delta_0 e^{i\phi(r)} \tag{1}$$

for which $\phi(r)$ is the electromagnetic gauge symmetry breaking macroscopic quantum phase. Heuristically, one may think of $\Delta_0 \equiv \sum_k \gamma \langle c_{k\uparrow} c_{-k\downarrow} \rangle$, where $c_{k\uparrow}$ ; $c_{-k\downarrow}$ are electron annihilation operators of opposite spin and momentum and $\gamma$ the pairing strength, as the amplitude of the electron-pair condensate wavefunction. A unidirectional PDW state is also a superconductor, but one that modulates the condensate order parameter spatially at wavevector $P_x$ such that

$$\Delta_x(\boldsymbol{r}) = \Delta_{P_x}(\boldsymbol{r})e^{iP_x x} + \Delta^*_{-P_x}(\boldsymbol{r})e^{-iP_x x} \tag{2}$$

In a tetragonal crystal, an orthogonal PDW state can also exist modulating with at wavevector $P_y$ along the *y*-direction as

$$\Delta_y(\boldsymbol{r}) = \Delta_{P_y}(\boldsymbol{r})e^{iP_y y} + \Delta^*_{-P_y}(\boldsymbol{r})e^{-iP_y y} \tag{3}$$

Because there are then five complex-valued scalar order parameter functions, when these three macroscopic quantum phases are intertwined, a plethora of global electron-pair order parameters becomes possible[1,2,3,10,11,12].

Theoretical analysis of how the DSC and two primary PDW states become intertwined requires an intricate Ginzburg-Landau-Wilson (GLW) free energy density functional[1,2,3,10]. Using such an approach, excellent success has been achieved in understanding the induced order parameters, the most prominent of which are the induced CDW states with order parameters[2,3] $\rho_{P_x,P_y}(\mathbf{r}) \propto \Delta_0^* \Delta_{P_x,P_y} + \Delta_{-P_x,-P_y}^* \Delta_0$ and $\rho_{2P_x,2P_y}(\mathbf{r}) \propto \Delta_{-P_x,-P_y}^* \Delta_{P_x,P_y}$. However, to fully understand the PDW of cuprates, one is not limited to studying such secondary or induced states, because SJTM imaging can give direct access[5-9] to the primary electron-pair orders. The structure and intertwining between PDW and DSC states are described by a subset of terms from the overall GLW free energy density

$$\mathcal{F} = \beta_{c1}|\Delta_0|^2 \left(\left|\Delta_{P_x}\right|^2 + \left|\Delta_{P_y}\right|^2 + \left|\Delta_{P_{-x}}\right|^2 + \left|\Delta_{P_{-y}}\right|^2\right) + \beta_{c2}\left[\Delta_0^2 \left(\Delta_{P_x}\Delta_{-P_x} + \Delta_{P_y}\Delta_{-P_y}\right)^* + (\Delta_0^2)^*\left(\Delta_{P_x}\Delta_{-P_x} + \Delta_{P_y}\Delta_{-P_y}\right)\right] \quad (4)$$

where only the lowest order coupling terms are considered and the gradient terms are ignored. Among the possible global electron-pair orders are[2,3] unidirectional or bidirectional modulated PDW phases $e^{iP_x x}$; $e^{iP_y y}$ with fixed amplitude, as in the Fulde-Ferrell[13] state; and unidirectional or bidirectional modulated PDW amplitudes $|\Delta_{P_x}|\cos(P_x x)$; $|\Delta_{P_y}|\cos(P_y y)$ with fixed phase, as in the Larkin-Ovchinnikov[14] state. An intriguing hypothetical state is a nematic pair density wave[2,3] with order parameter

$$N \equiv \left(\left|\Delta_{P_x}\right|^2 + \left|\Delta_{-P_x}\right|^2\right) - \left(\left|\Delta_{P_y}\right|^2 + \left|\Delta_{-P_y}\right|^2\right) \quad (5)$$

and such states may also exhibit vestigial nematic phases[15,16]. But, how the DSC and PDW orders are intertwined in cuprates, and their consequent global electron-pair order parameter, are all unknown.

Our objective is exploration of intertwined PDW and DSC states in cuprates, by using SJTM. In principle, the total electron-pair density at location $\mathbf{r}$, $n(\mathbf{r})$, can be visualized by measuring electron-pair (Josephson) critical-current $I_J(\mathbf{r})$ from the sample to a superconducting STM tip[17]. This is because $n(\mathbf{r}) \propto I_J^2(\mathbf{r})R_N^2(\mathbf{r})$ where $R_N$ is the normal-state junction resistance[18,19]. But since typical thermal fluctuations far exceed the Josephson energy $E_J$ between SJTM tip and sample surface, a phase-diffusive steady-state at voltage $V_J$ drives an electron-pair current $I_P(V_J) = \frac{1}{2}I_J^2 Z V_J/(V_J^2 + V_c^2)$. Here $V_c = 2eZk_BT/\hbar$ with $Z$ the high-frequency impedance of the junction[20,21]. In the theory of such spectra, the maximum value of electron-pair current $I_m = (\hbar/8ek_BT)I_J^2$, providing the basis for atomically resolved SJTM visualization of the electron-pair condensate in superconductors[5-9] as

$$n(\boldsymbol{r}) \propto I_\mathrm{m}(\boldsymbol{r})\, R_\mathrm{N}^2(\boldsymbol{r}) \qquad (6)$$

In that context, we study single crystals of $Bi_2Sr_2CaCu_2O_{8+x}$ with $CuO_2$ plaquette hole-density $p \approx 0.17$, by using a dilution refrigerator based SJTM[5]. The cryogenically cleaved samples terminate at the BiO crystal layer, and the *d*-wave superconducting scan-tip is prepared by exfoliating a nanometer scale $Bi_2Sr_2CaCu_2O_{8+\delta}$ flake from that sample surface[5,9]. A typical topographic image at T=45 mK then consists of atomically resolved and registered surface Bi atoms (Fig. 1A) with the typical measured $I_\mathrm{P}(V_J)$ shown as inset. Using such tips and a virtually constant $R_\mathrm{N}(\boldsymbol{r}) \approx 20$ MΩ as determined from analysis of topography at the setup voltage, we image $I_\mathrm{P}(V_J,\boldsymbol{r})$ and thus $I_\mathrm{m}(\boldsymbol{r})$, at T=45mK. The Fourier transform of the measured $I_\mathrm{m}(\boldsymbol{r})$ image, $I_\mathrm{m}(\boldsymbol{q})$, is shown in Fig. 1B. In such experiments on $Bi_2Sr_2CaCu_2O_{8+\delta}$, the $I_\mathrm{m}(\boldsymbol{r})$ and $I_\mathrm{m}(\boldsymbol{q})$ are both dominated by effects of the crystal *supermodulation*, a bulk 26Å periodic modulation of unit-cell dimensions [22] whose general effects are discussed elsewhere[9]. The focus of interest here is on the two PDWs observable in $I_\mathrm{m}(\boldsymbol{q})$ as four relatively broad peaks surrounding the wavevectors[5] $\boldsymbol{q} \approx \left(\frac{2\pi}{a}\right)(0,\pm 0.25); \left(\frac{2\pi}{a}\right)(\pm 0.25,0)$. Additionally, the somewhat heterogeneous electron-pair density of the DSC state is represented by the broad peak at $\boldsymbol{q} = \left(\frac{2\pi}{a}\right)(0,0)$. The phenomena detected separately at these five wavevectors can reveal key information on the interplay of the five order parameter functions of Eqns. 1-3.

To separately visualize the two PDWs we inverse Fourier transform $I_\mathrm{m}(\boldsymbol{q})$ in Fig. 1B for wavevectors $\pm \boldsymbol{P}_x \approx \left(\frac{2\pi}{a}\right)(0,\pm 0.25)$ and $\pm \boldsymbol{P}_y \approx \left(\frac{2\pi}{a}\right)(\pm 0.25,0)$ respectively, using the $\boldsymbol{q}$-space regions represented inside the colored circles. The resulting $I_{\mathrm{m}_x}(\boldsymbol{r})$, and $I_{\mathrm{m}_y}(\boldsymbol{r})$ images shown in Fig. 1C,D respectively, reveal the spatial arrangements of the two orthogonal PDW states (in the presence of a predominant DSC.) They are relatively coherent $4a_0$-periodioc modulations with amplitudes differently disordered, and without obvious correlation between them. Clearly, there is a proliferation of edge-dislocation $2\pi$ topological defects[23] in $I_{m_{x,y}}(r)$, whose ongoing study will be reported elsewhere. Here we concentrate on the interplay of between the simultaneously observed PDW and DSC states.

To do so, we parameterize the two PDW state signatures in terms

$$I_{m_{x,y}}(r) = A_{x,y}(r)\cos\left(P_{x,y}\cdot r + \delta_{x,y}(r)\right) \equiv A_{x,y}(r)\cos(\Phi_{x,y}(r)) \tag{7}$$

Fourier component selection at $\pm P_x = \left(\frac{2\pi}{a}\right)(0,\pm 0.25)$ and $\pm P_y = \left(\frac{2\pi}{a}\right)(\pm 0.25, 0)$ then yields

$$I'_m(r) == \frac{1}{\sqrt{2\pi}\sigma}\int dr' I_m(r')\, e^{-iP_{x,y}\cdot r'} e^{\frac{|r-r'|^2}{2\sigma_0^2}} \tag{8}$$

such that

$$A_{x,y}(r) \equiv 2\sqrt{\left(\mathcal{R}e\, I'_m(r)\right)^2 + \left(\mathcal{I}m\, I'_m(r)\right)^2} \tag{9}$$

(SI Section 1). The DSC state characteristic, $A_0(r)$, is defined similarly by inverse Fourier transform within the yellow circle surrounding wavevector $q = (0,0)$. Identical value of $\sigma_0 = \sigma_x = \sigma_y = 3\ nm$ are applied throughout. Using this approach, Figs. 2A,B,C respectively show $A_x(r)$, $A_y(r)$ and $A_0(r)$ visualized simultaneously in the FOV of Fig. 1A. Thus it becomes possible to directly explore the phenomenology of intertwining the DSC and PDW states (SI Section 2). Figure 2D presents measured $\langle A_x + A_y\rangle$ versus $A_0$, where the average is carried out over all locations $r$ at which $A_0(r)$ has the value indicated on the abscissa. The positive slope $s = 0.0389$ through zero indicates that, throughout the variations in $A_x(r)$, $A_y(r)$ and $A_0(r)$ (Figs. 1C,D and Figs. 2A,B,C), the PDW and the DSC states are mutually enhancing on the average. This conclusion is independent of any inadvertent heterogeneity in the normal-state Josephson junction resistance $R_N(r)$ in Eqn. 6, because it gets divided out by the definition of *s*. The implication of Fig. 2 it that, within the GLW context (Eqn. 4 and Refs. 1,2,3) the DSC and PDW states are attractive $((\beta_{c1} - |\beta_{c2}|) < 0)$ electronic phases (SI Section 4) at zero magnetic field, although there is evidence of repulsion when strong gradients present as in the vortex core.

Next, to search for the hypothetical[2,3] nematic PDW state we define a nematic order parameter

$$\mathcal{N}(\mathbf{r}) = \{A_x(\mathbf{r}) - A_y(\mathbf{r})\}/\{A_x(\mathbf{r}) + A_y(\mathbf{r})\} \tag{10}$$

Analyzing the data from Figs. 2A,B in this way generates $\mathcal{N}(\mathbf{r})$ as shown in Fig. 3A. Again we note that Eqn. 10 provides an empirical definition for a nematic PDW state based directly on $I_m(\mathbf{r})$ data, i.e. independent of any possible variations in $R_N(\mathbf{r})$. Figure 3A then reveals the existence of strong PDW nematicity in Bi$_2$Sr$_2$CaCu$_2$O$_8$. Indeed, the histogram of all $\mathcal{N}(\mathbf{r})$ values from Fig. 3A in the inset demonstrates that $|\mathcal{N}(\mathbf{r})| > 0.3$ for 45% of the sample area and thus that these nematic PDW states can predominate. To exemplify, Fig. 3B shows examples of measured $I_m(\mathbf{r})$ along the x-axis (y-axis) in its left(right) panels, both within domains where $\mathcal{N}(\mathbf{r}) \gg 0$. Figure 3C shows equivalent exemplary data for domains where $\mathcal{N}(\mathrm{r}) \ll 0$. Overall, we find that a robust nematic PDW state, consisting of nanoscale domains of opposite nematicity in electron-pair density, occurs in Bi$_2$Sr$_2$CaCu$_2$O$_8$ at $p \approx 0.17$.

But this discovery begs the question of what it is that sets the size and location of the nematic PDW domains? One clue comes from recent reports[24] that, in La$_{2-x}$Ba$_x$CuO$_4$ away from $p \approx 0.125$, introducing Zn atom substitution at 1% of Cu sites, leads to a cascade transition from 2D superconductivity to 3D superconductivity with falling temperature. The inference derived from these studies is that the Zn atoms pins PDW order locally. In cuprates, at each Zn impurity atom there is a superconductive impurity state with energy $E \approx -1 meV$, and also a powerful suppression of electron-pair condensate $n(\mathbf{r})$ as exemplified by Fig. S2F. In Bi$_2$Sr$_2$CaCu$_2$O$_8$ the Zn impurity states can be imaged directly in superconducting-tip differential conductance imaging, by finding the local maxima in $Z(\mathbf{r}) \equiv I(20\mathrm{mV}, \mathbf{r}) - I(-20\mathrm{mV}, \mathbf{r})$ where I(V) is the superconductor-insulator-superconductor (SIS) single-electron tunnel current. These maxima occur because the coherence peaks of SJTM tip density-of-states near $E \approx \pm 20 meV$ are convoluted with the Zn impurity state density-of-states peak $E \approx -1 meV$ to produce a strong jump in tunnel currents $I(E \approx \pm 20 meV, \mathbf{r})$. Figure 4A shows the resulting image $Z(\mathbf{r}) \equiv I(20\mathrm{mV}, \mathbf{r}) - I(-20\mathrm{mV}, \mathbf{r})$ from which each Zn impurity-state maximum is identified by blue circles (SI Section 3). In Figs. 4B we show the amplitude of nematic order parameter $|\mathcal{N}(\mathbf{r})|$ with the sites of Zn impurity resonances overlaid as blue dots. Visually the Zn sites seem to occur near the domain boundaries in $|\mathcal{N}(\mathbf{r})|$. This can be quantified by plotting the histogram of the distribution of distances between each Zn impurity atoms and its nearest PDW domain wall and comparing to the expected average distance of uncorrelated random points. The result shown in Fig. 4C reveals that the Zn sites occur highly preferentially near the boundaries of the PDW Ising domains. This direct visualization indicates that the PDW nematic domains are pinned by interactions with Zn impurity atoms at the Cu sites, and is therefore highly consistent with deductions from the transport studies of Zn-doped $La_{2-x}Ba_xCuO_4$.[24]

Spectroscopic imaging STM and resonant X-ray scattering have produced a wealth of understanding of unidirectionality, commensuration and domain formation for CDW modulations in strongly underdoped cuprates. However, because photon scattering cannot (yet) detect PDW states, because the pseudogap in single particle tunneling masks the true electron-pairing energy gap and thus order parameter, and because SJTM visualization of PDW states is recent[5-9], equivalent issues for the cuprate PDW state are unresolved. Our SJTM visualization now demonstrates that the $Bi_2Sr_2CaCu_2O_8$ PDW states tend strongly to be both unidirectional and commensurate, a situation widely predicted[25-34] as a consequence of strong-coupling physics within the $CuO_2$ Hubbard model. Further, we note that visualization of these characteristics in a robust PDW state occurs in $Bi_2Sr_2CaCu_2O_{8+x}$ at a carrier density where the charge density wave (CDW) modulations are virtually nonexistent[35,36], as is the empirical case in the samples studied here (SI Fig. S3). The implication, consistent with the eight unit-cell periodic energy gap modulations observed in single particle tunneling[37,38], is that the $Bi_2Sr_2CaCu_2O_8$ PDW states are not induced by the existence of a CDW and instead are the primary translation symmetry breaking state of hole-doped $CuO_2$. Furthermore, since the spatial configurations of the PDWs studied here appear uninfluenced by a preexistent CDW yet are obviously disordered, effects of chemical (and possibly dopant-ion) randomness on the PDW are adumbrated. Indeed, recent transport studies[24] provide experimental evidence that Zn impurity atoms pin the PDW order in $La_{2-x}Ba_xCuO_4$. For comparison, our visualization of the $Bi_2Sr_2CaCu_2O_8$ PDW nematic domains simultaneously with Zn scattering resonances demonstrates directly that the Zn impurity-atom sites occur predominantly within boundary regions between these domains. Overall, a vestigial state is one that only partially breaks the symmetry of the true ordered state. Here we find a nematic

PDW state breaking the rotational ($C_4$) symmetry of the lattice but not long-range translational symmetry, in what appears to be a disorder-pinned realization of a global nematic order. Hence, a plausible context in which to consider the PDW in $Bi_2Sr_2CaCu_2O_{8+x}$ is as a new form of vestigial nematic state based on a disordered unidirectional density wave of electron pairs[39] rather than of single electrons[40].

Ergo, by using SJTM to simultaneously visualize the PDW and DSC states of $Bi_2Sr_2CaCu_2O_8$ near $p \approx 0.17$ where no CDW phenomena obtrude, we demonstrate that they are intertwined in mutually attractive phases (Fig. 2). Conversely, separate imaging of the electron-pair density modulations of the two orthogonal PDWs reveals a robust nematic PDW state, with primarily unidirectional and lattice-commensurate electron-pair density modulations forming Ising domains of opposite nematicity (Fig. 3). Imaging the sites of Zn impurity-states finds them occurring preferentially in boundary regions of minimal nematic order (Fig. 4) implying that the PDW domains are pinned thereby. Generally, these data signify that the PDW state of $Bi_2Sr_2CaCu_2O_{8+x}$ is a vestigial nematic phase of electron pairs, that it is locally unidirectional and lattice commensurate, and that it is not a subordinate but a primary electronic order of hole-doped $CuO_2$.

**FIGURES**

**Figure 1: Visualizing Electron Pair Density $n(r)$**

A. SJTM topographic image $T(r)$ of BiO termination layer of Bi$_2$Sr$_2$CaCu$_2$O$_8$. Inset shows average electron-pair current spectrum $I_P(V_J)$ measure in this FOV at T=45mK and $R_N \approx 20$ MOhm, with maxima occurring at $\pm I_m$.

B. Power spectral density Fourier transform of $I_m(r)$, $I_m(\mathbf{q})$, as measured in FOV of 1a. Four broad PDW peaks surround the wavevectors $\mathbf{P} = (2\pi/a)(0, \pm 0.25); (2\pi/a)(\pm 0.25, 0)$ as indicated by pairs of red and blue arrows respectively. The DSC electron-pair density is represented by the broad peak at $\mathbf{q} = (2\pi/a)(0,0)$ as indicated by the yellow circle.

C. Fourier filtration of $I_m(\mathbf{q})$ $\mathbf{P} = (2\pi/a)(0, \pm 0.25)$ as in Fig. 1B, to visualize the $\pm \mathbf{P}_x$ PDW modulating along the CuO$_2$ x-axis.

D. Fourier filtration of $I_m(\mathbf{q})$ $\mathbf{P} = (2\pi/a)(\pm 0.25, 0)$ in Fig. 1B, to visualize the $\pm \mathbf{P}_y$ PDW modulating along the CuO$_2$ y-axis.

**Figure 2: Intertwined DSC and PDW Order Parameters**

A. Amplitude of $I_m(\mathbf{r})$ modulations $A_{P_x}(\mathbf{r})$ for PDW state with $\pm \mathbf{P}_x$, from 1C.

B. Amplitude of $I_m(\mathbf{r})$ modulations $A_{P_y}(\mathbf{r})$ for PDW state with $\pm \mathbf{P}_y$, from 1D.

C. Amplitude of DSC electron-pair density $A_0(\mathbf{r})$ derived from Eqn. 6 with *q*=0 as indicated by yellow circle in Fig. 1B.

D. $\langle A_x(\mathbf{r}) + A_y(\mathbf{r})\rangle$ averaged over all locations *r* where $A_0(\mathbf{r})$ equals the abscissa value $A_0$. The solid line is a linear fit to these data through (0,0). Inset shows the 2D histogram of same data.

**Figure 3: Nematic PDW State of Bi$_2$Sr$_2$CaCu$_2$O$_8$**

A. Nematic order parameter $\mathcal{N}(\mathbf{r}) = \{A_x(\mathbf{r}) - A_y(\mathbf{r})\}/\{A_x(\mathbf{r}) + A_y(\mathbf{r})\}$ derived from 2A,B. Domains of opposite nematicity occur with correlation length $\xi \approx 15$nm. `Inset: Histogram of all $\mathcal{N}(\mathbf{r})$ values in A, showing nonzero mean value. The magnitude |

$\mathcal{N}| > 0.3 |$ for approximately 45% of the FOV, indicating a strong nematic interaction between the two PDW.

B. At left, four examples of measured $I_m(\mathbf{r})$ along the x-axis; at right, four examples of measured $I_m(\mathbf{r})$ along the y-axis : both within domains where $\mathcal{N}(\mathbf{r}) \gg 0$.

C. At left, four examples of measured $I_m(\mathbf{r})$ along the x-axis; at right, four examples of measured $I_m(\mathbf{r})$ along the y-axis : both within domains where $\mathcal{N}(\mathbf{r}) \ll 0$

**Figure 4: Pinning of PDW Nematic Domains by Zn Impurity Atoms**

A. Locations of zinc impurity atoms Z(**r**) in A, as detected in $Z(r) \equiv I\ (r, 20\text{mV}) - I\ (r, -20\text{mV})$ are shown as blue circles. The identification and register of Zn impurity sites is discussed in SI section 3.

B. $|\mathcal{N}(r)|$, the amplitude of the nematic order parameter from Fig. 3A, with the sites of Zn impurity resonances overlaid as blue dots.

C. The distribution of distances between each Zn impurity atoms and its nearest PDW domain walls (red). This is compared to the expected average distance if there is no correlation between Zn impurity atoms and the PDW domain walls(blue). Zn impurity atoms are concentrated near the PDW domain walls.

**Acknowledgements**: We acknowledge and thank Owen H.S. Davis and Shuqiu Wang for key discussions and suggestions. M.H.H. and J.C.S.D. acknowledge support from the Moore Foundation's EPiQS Initiative through Grant GBMF9457. W.R. and J.C.S.D. acknowledge support from the European Research Council (ERC) under Award DLV-788932. W.C. and J.C.S.D. acknowledge support from the Royal Society under Award R64897. N.K., S.O'M and J.C.S.D. acknowledge support from Science Foundation of Ireland under Award SFI 17/RP/5445. H.E. acknowledges support from JSPS KAKENHI (No. JP19H05823). PDJ acknowledges support by QuantEmX grant GBMF9616 from ICAM / Moore Foundation and by a Visiting Fellowship at Wadham College, Oxford., UK

**Author Contributions:** W.C. and J.C.S.D. conceived the project. P.D.J. and J.C.S.D. supervised the research. H.E. and S.U. synthesized and characterized the samples; M.H.H. and J.C.S.D. carried out SJTM measurements. Analysis and scholarship was carried out by W.C., W.R., N.K., P.D.J. and S.O'M. J.C.S.D. and W.C wrote the paper with key contributions from N.K., W.R. and P.D.J.. The manuscript reflects the contributions and ideas of all authors.

**Author Information** Correspondence and requests for materials should be addressed to [jcseamusdavis@gmail.com](mailto:jcseamusdavis@gmail.com).

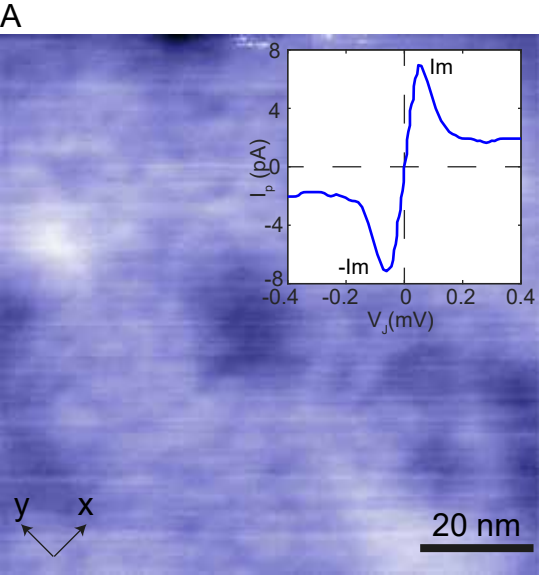
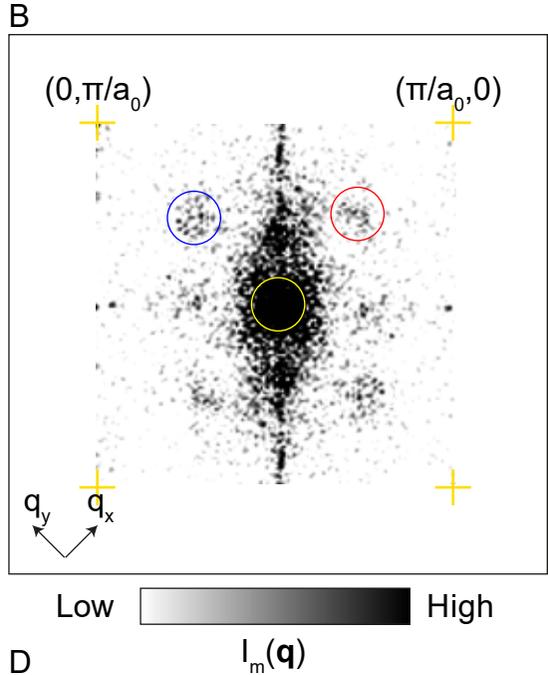
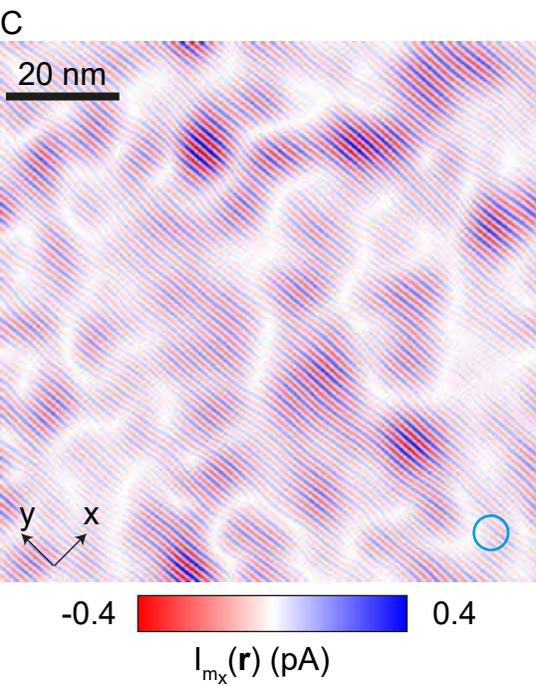
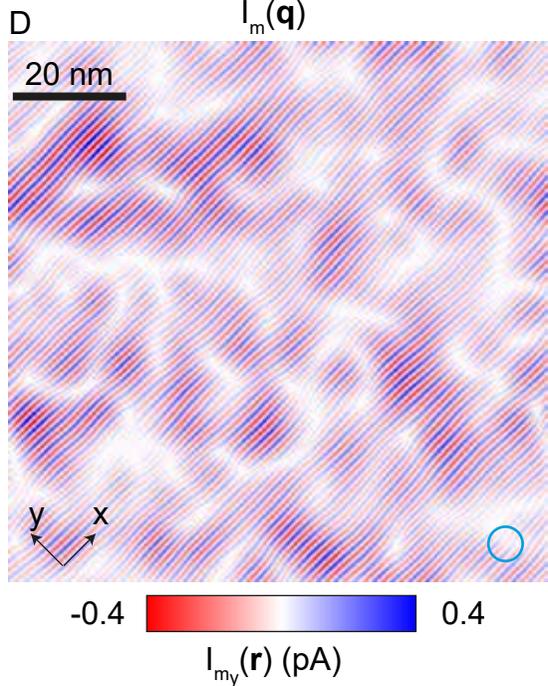

**Figure 1**

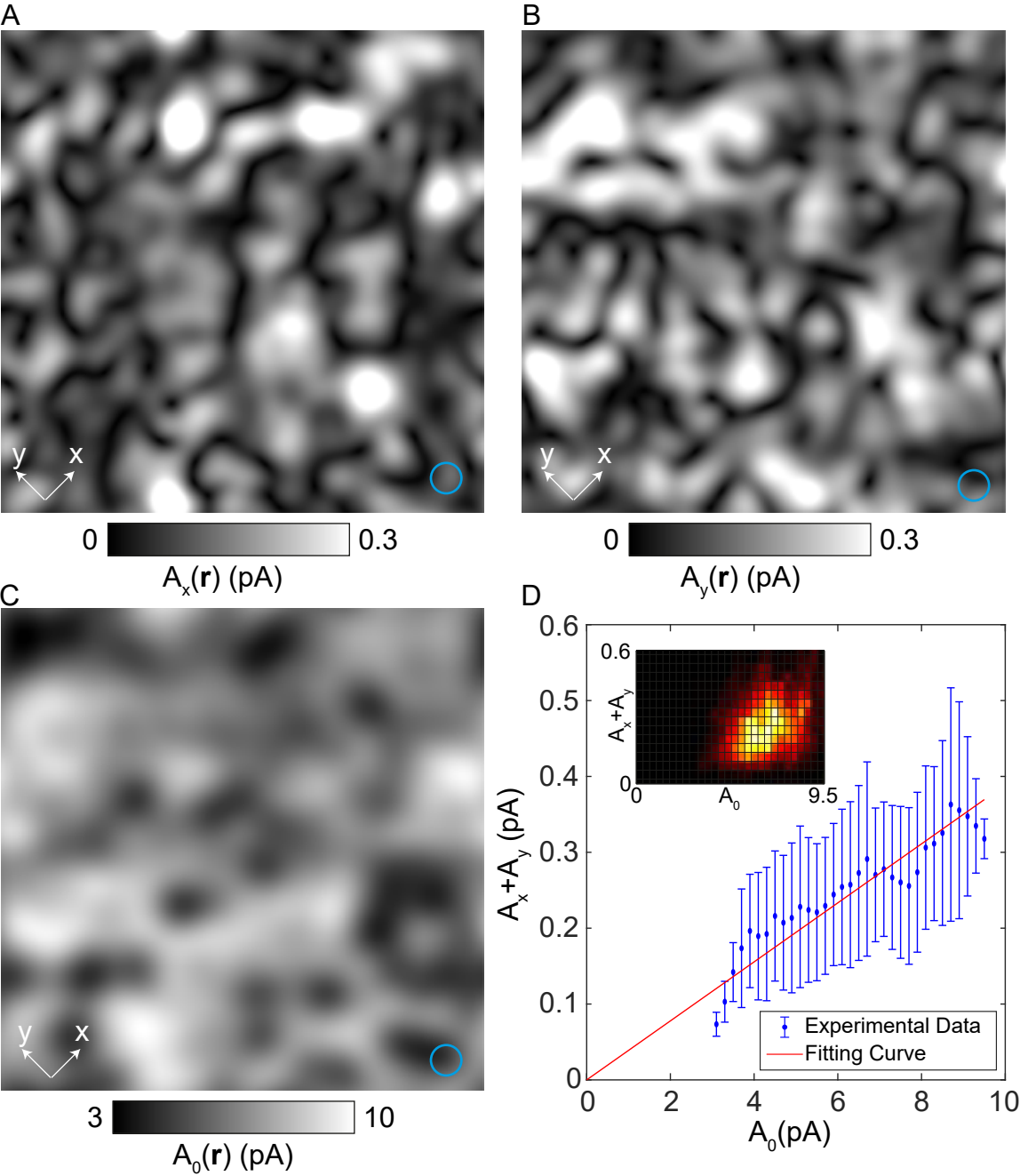

**Figure 2**

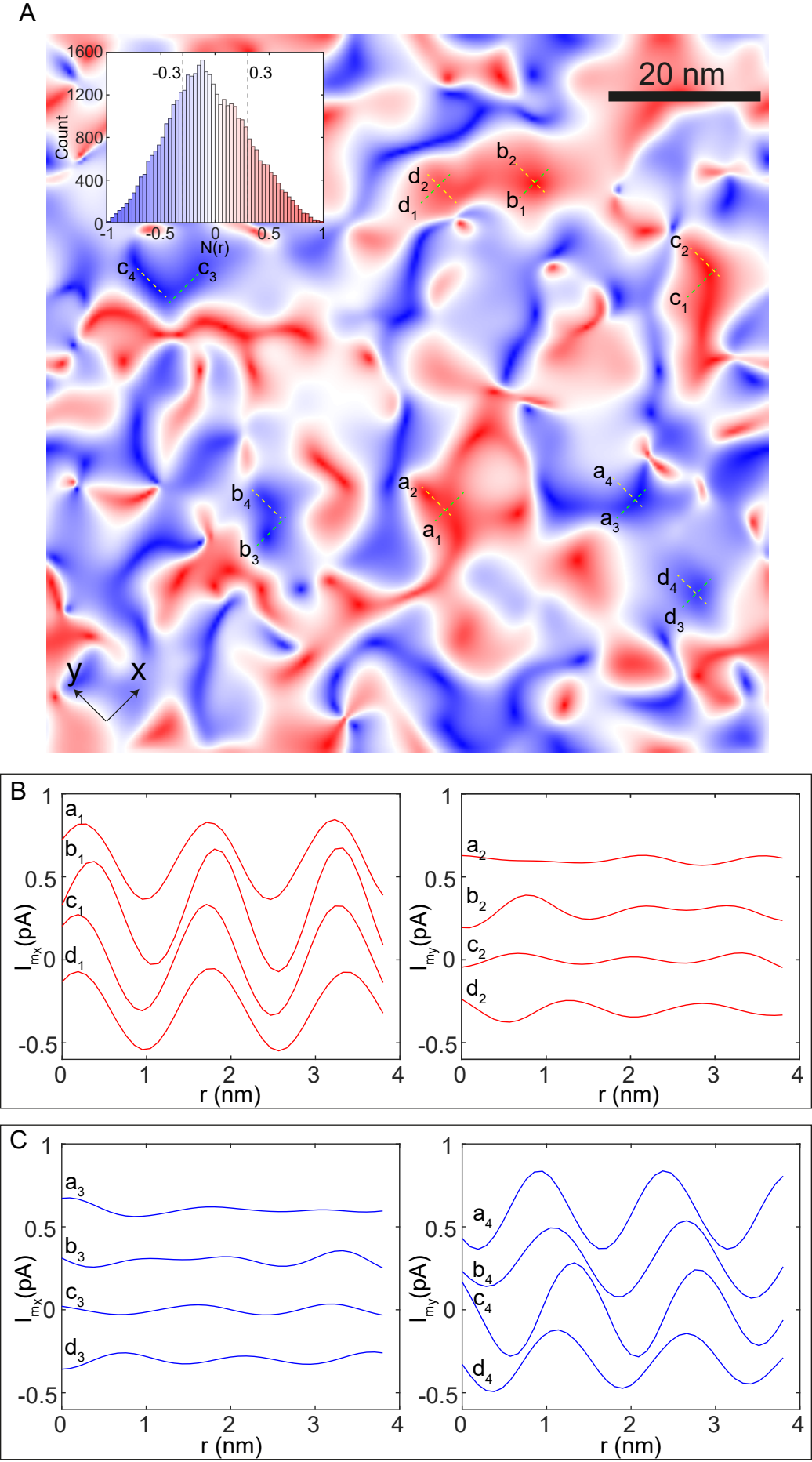

Figure 3

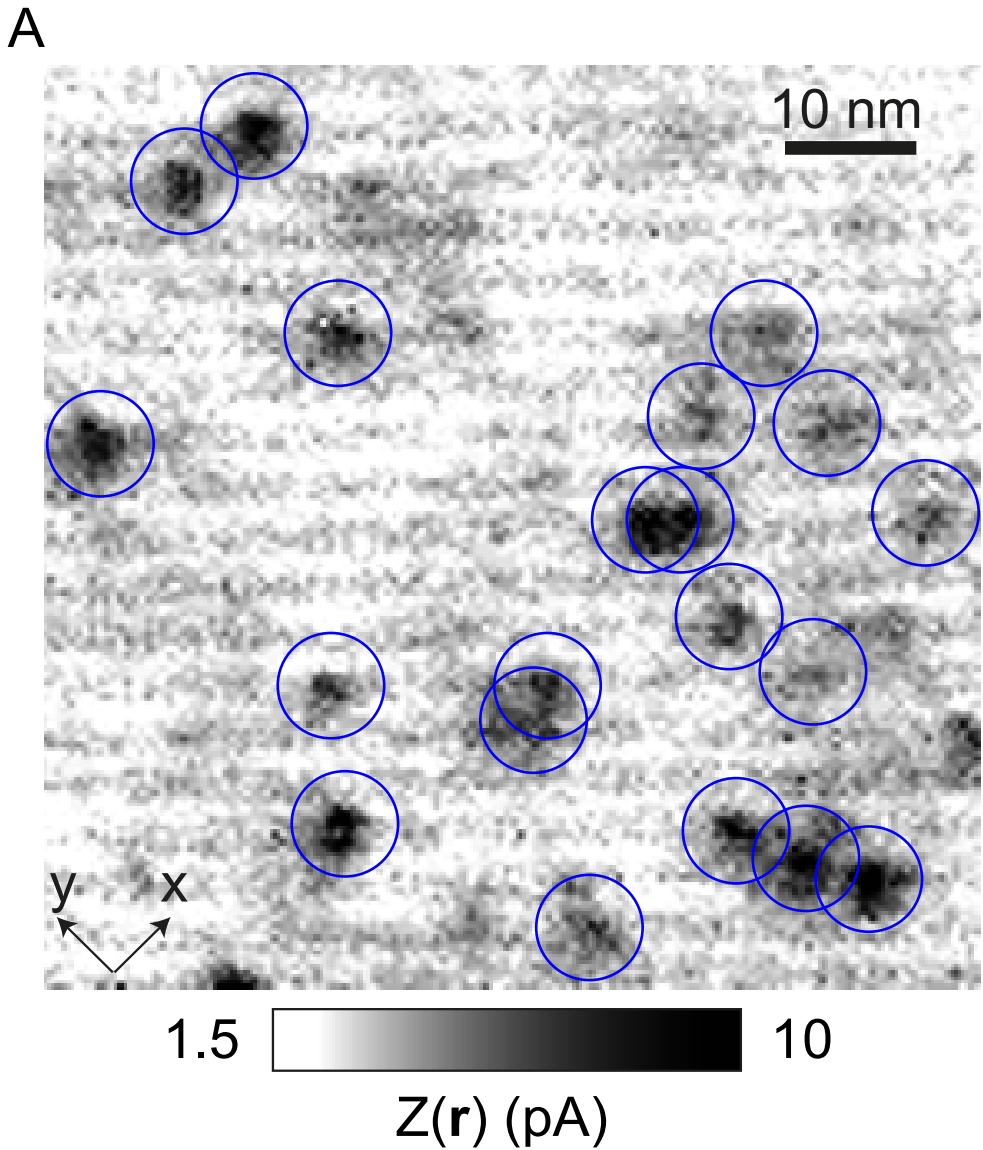

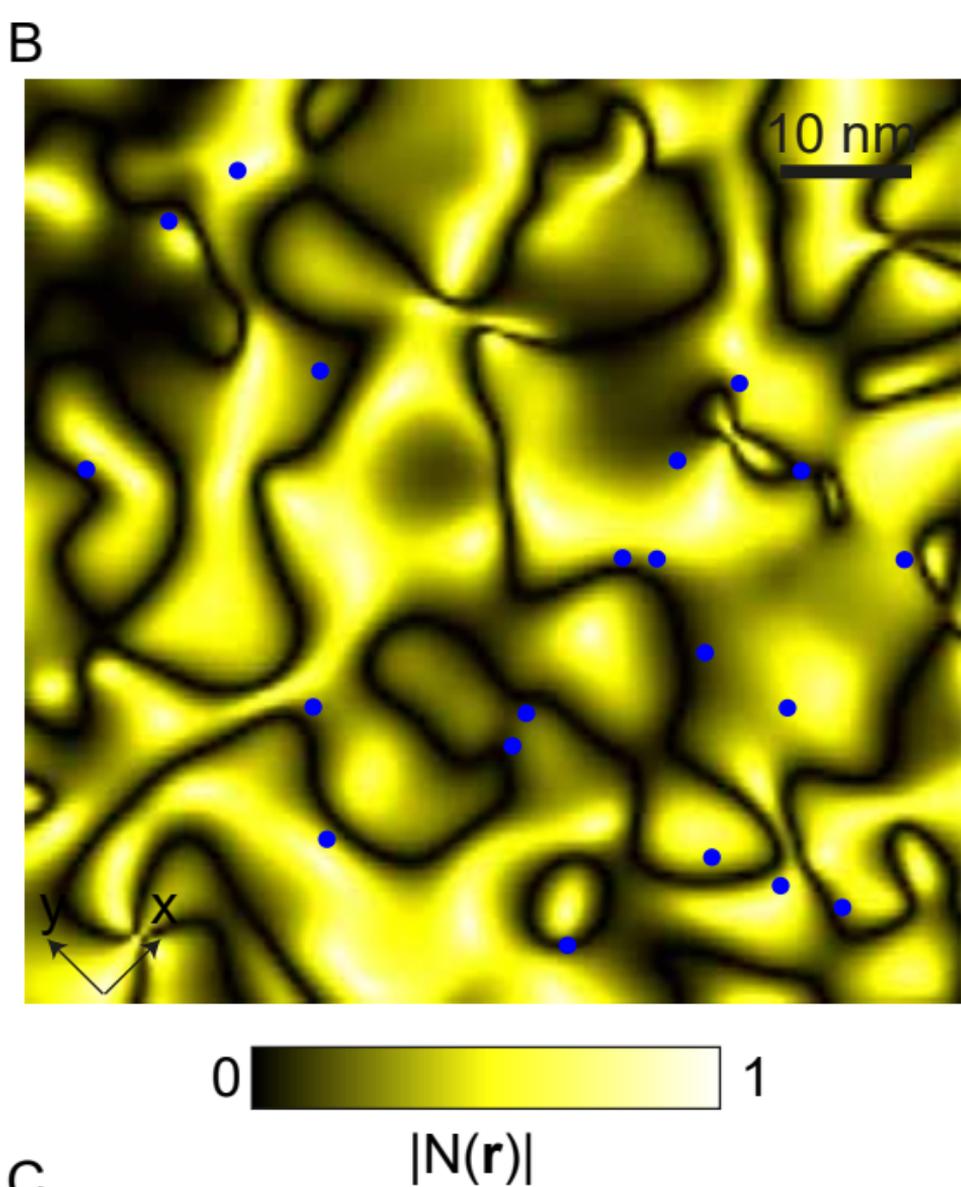

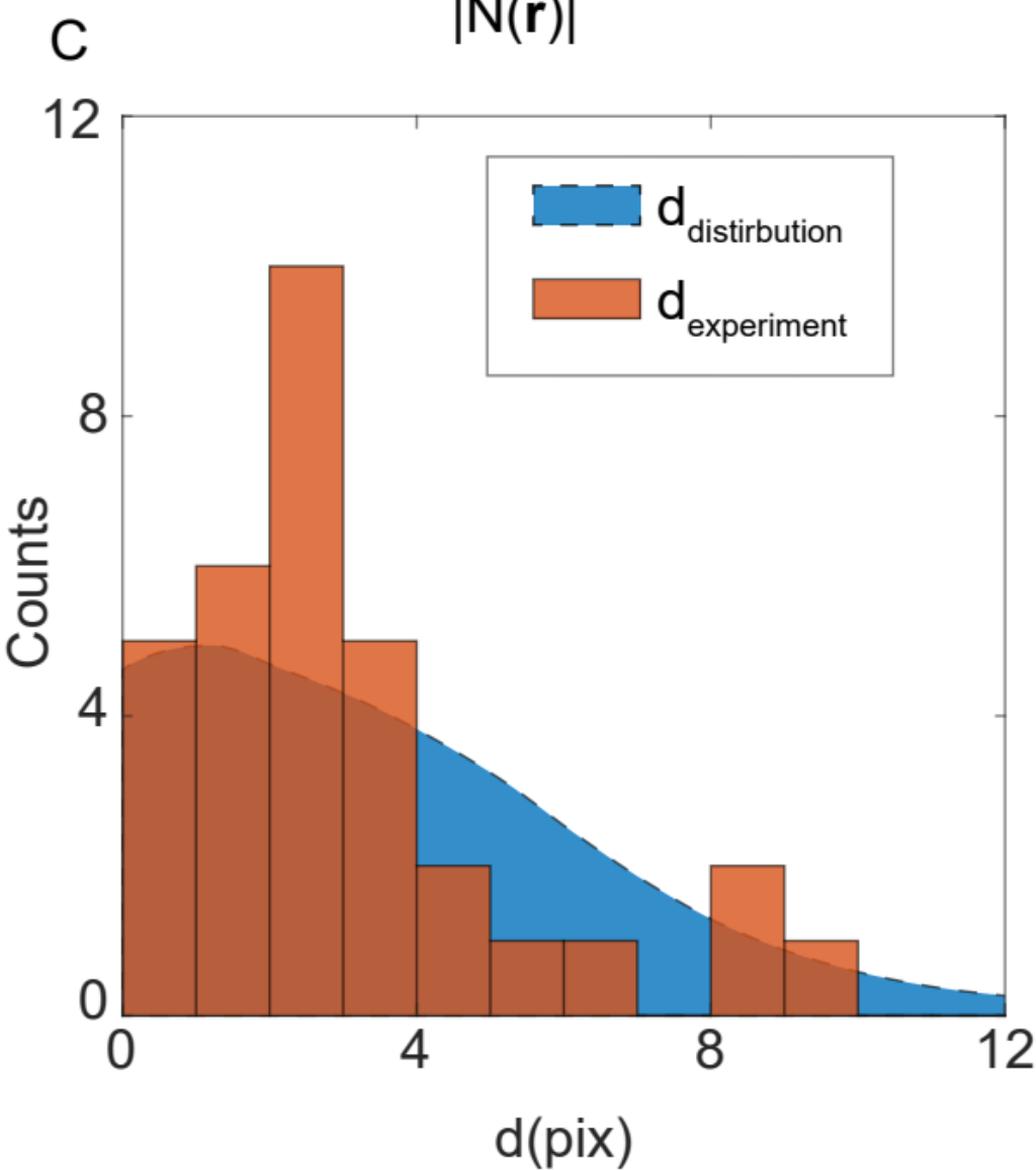

**Figure 4**

*Supporting Information for*

# Identification of a Nematic Pair Density Wave State in Bi$_2$Sr$_2$CaCu$_2$O$_{8+x}$

Weijiong Chen, Wangping Ren, Niall Kennedy, M. H. Hamidian, S. Uchida, H. Eisaki, Peter. D. Johnson, Shane M. O'Mahony and J.C. Séamus Davis

## 1. Fourier Decomposition of SJTM Data

To separate the components of PDW ($q = P_{x,y}$) in the $I_m(\mathbf{r})$ map, we use a two-dimensional lock-in method. We parameterize the two PDW state signatures in terms

$$I_{m_{x,y}}(r) = A_{x,y}(r)\cos\left(P_{x,y}\cdot r + \delta_{x,y}(r)\right) \equiv A_{x,y}(r)\cos(\Phi_{x,y}(r)) \quad (1)$$

Fourier component selection at $\pm P_x \approx \left(\frac{2\pi}{a}\right)(0,\pm 0.25)$ and $\pm P_y \approx \left(\frac{2\pi}{a}\right)(\pm 0.25, 0)$. $A_{x,y}(r)$ and $\delta_{x,y}(r)$ are the local amplitude and phase shift of each PDW component, respectively. To separate these signatures from the $I_m(\mathbf{r})$ map, we carry out a two-dimensional lock-in method in which references $\alpha(r) = \cos(P_{x,y}\cdot r)$ and $\beta(r) = \sin(P_{x,y}\cdot r)$ are multiplied by $I_m(\mathbf{r})$

$$X(r) = I_m(\mathbf{r})\cdot \alpha(r)$$

$$Y(r) = I_m(\mathbf{r})\cdot \beta(r) \quad (2)$$

In practice, it is more convenient to calculate $X(r)$ and $Y(r)$ together in complex number with $\alpha(r) - i\cdot\beta(r) = e^{-iP_{x,y}\cdot r}$. Then, low pass filtering these products:

$$I_m'(r) = = \frac{1}{\sqrt{2\pi}\sigma}\int dr' I_m(r') e^{-iP_{x,y}\cdot r'} e^{\frac{|r-r'|^2}{2\sigma^2}} \quad (3)$$

We use the filter width $\sigma = 3\ nm$. The real part of $I_m'(r)$ contains $A_{x,y}(r)$ and $\cos\left(\delta_{x,y}(r)\right)$. And the imaginary part of $I_m'(r)$ contains $A_{x,y}(r)$ and $\sin\left(\delta_{x,y}(r)\right)$

$$\mathcal{R}e\ I_m'(r) = \frac{A_{x,y}(r)\cos\left(\delta_{x,y}(r)\right)}{2} \quad (4)$$

$$\mathcal{I}m\, I_m{}'(\boldsymbol{r}) = \frac{A_{x,y}(\boldsymbol{r})\sin\left(\delta_{x,y}(\boldsymbol{r})\right)}{2} \tag{5}$$

Finally, we achieve the amplitude and phase shift of $\boldsymbol{P}_{x,y}$ PDW state by equation

$$A_{x,y}(\boldsymbol{r}) \equiv 2\sqrt{\left(\mathcal{R}e\, I_m{}'(\boldsymbol{r})\right)^2 + \left(\mathcal{I}m\, I_m{}'(\boldsymbol{r})\right)^2} \tag{6}$$

$$\delta_{x,y}(\boldsymbol{r}) = \arctan\left(\frac{\mathcal{I}m\, I_m{}'(\boldsymbol{r})}{\mathcal{R}e\, I_m{}'(\boldsymbol{r})}\right) \tag{7}$$

$A_0(\boldsymbol{r})$ (fig. 2C) is calculated by the same algorithm with q=0, which is equivalent to low pass filtering $I_m(\mathbf{r})$ map with $\sigma = 3\, nm$.

## 2. Bi$_2$Sr$_2$CaCu$_2$O$_{8+\delta}$ tip segregation of Pair Tunneling to DSC and PDW Orders

Here, we discuss the link between the $P_{x,y} \approx (0, \pm 0.25)\, and\, (\pm 0.25, 0)\frac{2\pi}{a_0}$ observed in electron pair density modulations in the present study and the $Q_{x,y} \approx (0, \pm 0.125)\, and\, (\pm 0.125, 0)\frac{2\pi}{a_0}$ as reported in energy gap Δ(r) imaging of Refs. [1],[2]. It has been proposed that this distinction in wavevectors is due to the segregation of pair tunneling to DSC and PDW orders when using a Bi$_2$Sr$_2$CaCu$_2$O$_{8+\delta}$ nanoflake tip.

A heuristic model for pair tunneling from a Bi$_2$Sr$_2$CaCu$_2$O$_{8+\delta}$ nanoflake tip to a Bi$_2$Sr$_2$CaCu$_2$O$_{8+\delta}$ bulk crystal, both of which have coexisting DSC order and PDW order parallel to the sample surface is shown in Fig. S1. The perfect pair tunneling can only occur between the states with the same momentum of the center of mass because of the conservation of momentum. Such that, the $c_k^\dagger c_{-k}^\dagger$ pairs of the homogenous DSC state at one side can only tunnel into the same DSC pairing states at another side. And, the $c_k^\dagger c_{-k+Q}^\dagger$ pair of the modulating PDW state at one side can only tunnel into the same PDW pairing states at another side. These are marked as orange and blue arrows in Fig. S1.

Considering the pair tunneling in the PDW channel, the tip's PDW order parameter $\Psi_T$ and the sample's PDW order parameter $\Psi_S$ are written as

$$\Psi_T = \Delta_1 e^{i\theta_1} \exp[iQ(x+\delta)] \text{ and } \Psi_S = \Delta_2 e^{i\theta_2} \exp(iQx) \tag{8}$$

where x is the position and $\theta$ is the phase of the order parameter. In this case, the inter-PDW Josephson current takes the form

$$I_J^P \propto \Psi_T \Psi_S^* - \Psi_T^* \Psi_S \propto \sin(Q\delta + \phi) \tag{9}$$

Our present study using SJTM measures the local cooper pair density $n(r) \propto I_J^2$. If the Josephson tunneling of PDW and DSC channels are independent, then,

$$n(r) \propto I_J^{S^2} + I_J^{P^2} \tag{10}$$

The pair density of DSC part $I_J^{S^2}$ is locally approximately constant while the pair density of PDW part modulates in the form:

$$I_J^{P^2} \propto \sin^2(Q\delta + \phi) \tag{11}$$

Here $\delta$ is the transverse displacement between the $Bi_2Sr_2CaCu_2O_{8+\delta}$ nanoflake tip and the $Bi_2Sr_2CaCu_2O_{8+\delta}$ sample. Therefore in such models[2], if the modulation of the order parameter has a wave vector Q≈ $(0,0.125)\frac{2\pi}{a_0}$ or wave length $\lambda \approx 8a_0$, the modulation of our measured pair density would have a primary periodicity of $4a_0$.

### 3. Identification and Register of Zn Impurity Sites

In $Bi_2Sr_2CaCu_2O_{8+\delta}$, the scanning tunneling spectrum on the Zn site shows the intense scattering resonance peak centered at $E \approx -1\ meV$ (3). This resonance peak is detected at higher energy (Fig. S2A) in our experiments, due to the superconductor-insulator-superconductor (SIS) tunnelling junction. The superconducting tip with coherence peaks near $E \approx \pm 20 meV$ results in the convoluted tunnelling current I shows a strong jump near $E \approx \pm 20 meV$, comparing to the I-V curve away from the Zn sites (Fig. S2B).

Therefore, we can identify the Zn impurities by finding the local maxima in $Z(r) \equiv I(20mV, r) - I(-20mV, r)$ map. Fig. S2E shows the $Z(r)$ map with identified Zn impurities marked by blue circles.

Because $Z(\mathbf{r})$ map and $I_m(r)$ maps are measured using independent experiments but in the same FOV, there is usually a small displacement of two sets of data. So, the positions of these Zn impurities in $Z(\mathbf{r})$ map must be transformed to their correct positions in $I_m(r)$ map. We establish a transformation matrix. We firstly manually choose 3 Zn impurities in both $Z(\mathbf{r})$ map ($a_1 = (x_{a_1}, y_{a_1})^T, a_2 = (x_{a_2}, y_{a_2})^T, a_3 = (x_{a_3}, y_{a_3})^T$) (blue circles in Fig. S2C) and $I_m(r)$ map $b_1 = (x_{b_1}, y_{b_1})^T, b_2 = (x_{b_2}, y_{b_2})^T, b_3 = (x_{b_3}, y_{b_3})^T$ (cyan crosses in Fig. S2D). Then, we generate the transformation matrix M by equation:

$$M = [b_2 - b_1 \quad b_3 - b_1] \times [a_2 - a_1 \quad a_3 - a_1]^{-1} \tag{12}$$

Thus, finally, we can transform the position of each Zn impurity in Z(r) map $a_{Zn}$ to its position in Im(r) map $b_{Zn}$ by

$$b_{Zn} = M \times (a_{Zn} - a_1) + b_1 \tag{13}$$

Fig. S2F shows the final Zn impurities in $I_m(r)$ map which are transformed from Zn impurities in Z(r) map (Fig. S2E).

### 4. Intertwining PDW with DSC orders in GLW free energy density

The structure and intertwining between the PDW orders $(\Delta_{P_x}(\mathbf{r}), \Delta_{P_y}(\mathbf{r}), \Delta_{-P_x}(\mathbf{r}), \Delta_{P_y}(\mathbf{r}))$ and DSC order ($\Delta_0$) is described by a subset of terms from the overall GLW free energy density, as mentioned in the main text and ref. 4,5,6.

$$\mathcal{F} = \beta_{c1}|\Delta_0|^2 \left( |\Delta_{P_x}|^2 + |\Delta_{P_y}|^2 + |\Delta_{P_{-x}}|^2 + |\Delta_{P_{-y}}|^2 \right)$$
$$+\beta_{c2} \left[ \Delta_0^2 \left( \Delta_{P_x}\Delta_{-P_x} + \Delta_{P_y}\Delta_{-P_y} \right)^* + (\Delta_0^2)^* \left( \Delta_{P_x}\Delta_{-P_x} + \Delta_{P_y}\Delta_{-P_y} \right) \right] \tag{14}$$

To minimize the free energy and find the ground state, one must insert the phase terms of order parameters.

$$\Delta_0 = |\Delta_0|e^{i\phi_0}, \Delta_{P_x}(\mathbf{r}) = |\Delta_x|e^{i\phi_1}, \Delta_{P_y}(\mathbf{r}) = |\Delta_y|e^{i\phi_2},$$
$$\Delta_{-P_x}(\mathbf{r}) = |\Delta_x|e^{i\phi_3}, \Delta_{P_y}(\mathbf{r}) = |\Delta_y|e^{i\phi_4} \tag{15}$$

Here, for simplicity, we ignore the case of the loop current order. Such that

$$\begin{aligned}\mathcal{F} = &\ 2|\Delta_0|^2|\Delta_x|^2[\beta_{c1} + \beta_{c2}\cos(2\phi_0 - \phi_1 - \phi_3)] \\ &+ 2|\Delta_0|^2|\Delta_y|^2[\beta_{c1} + \beta_{c2}\cos(2\phi_0 - \phi_2 - \phi_4)]\end{aligned} \quad (16)$$

The coupling term $\beta_{c2}$ locks the phase of the uniform DSC order the PDW orders[6]. Obviously, the minimum free energy is:

$$\mathcal{F} = 2|\Delta_0|^2\left(|\Delta_x|^2 + |\Delta_y|^2\right)[\beta_{c1} - |\beta_{c2}|] \quad (17)$$

Therefore, the positive correlation between $|\Delta_0|^2$ and $|\Delta_x|^2 + |\Delta_y|^2$ in Fig. 2D implies $\beta_{c1} - |\beta_{c2}|<0$ under the zero magnetic field circumstances of that experiment.

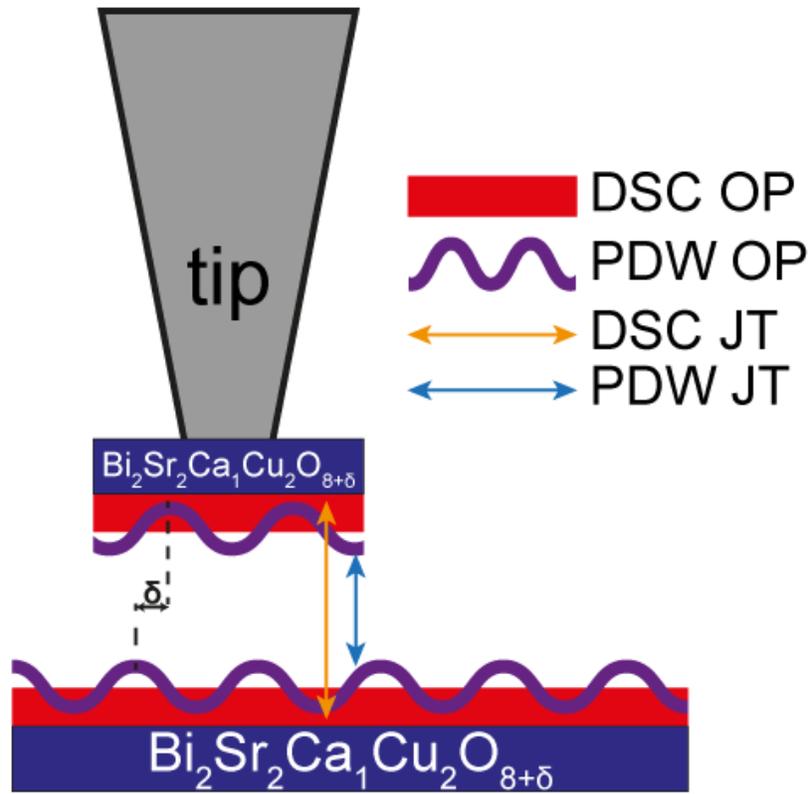

Figure S1. Schematic of possible Josephson tunneling (JT) in the presence of d-wave superconductivity (DSC) and pair density wave (PDW) order parameters (OPs).

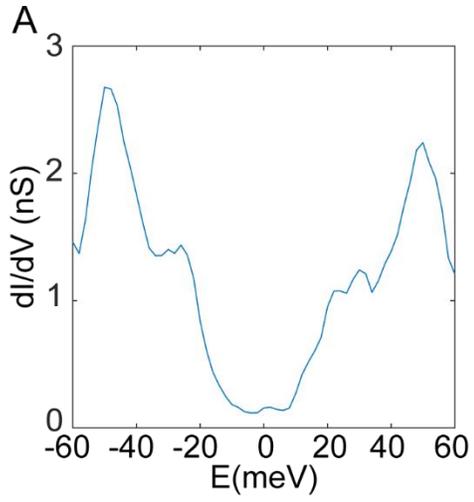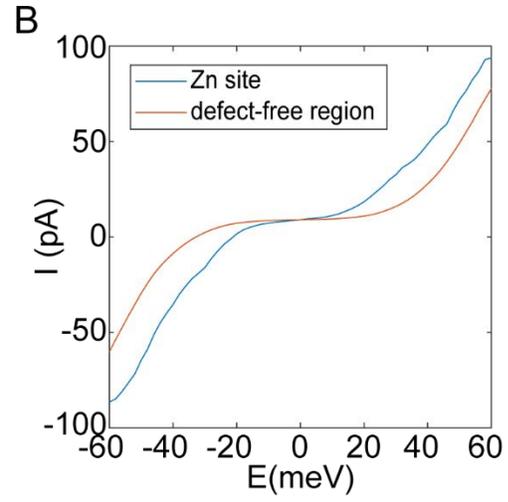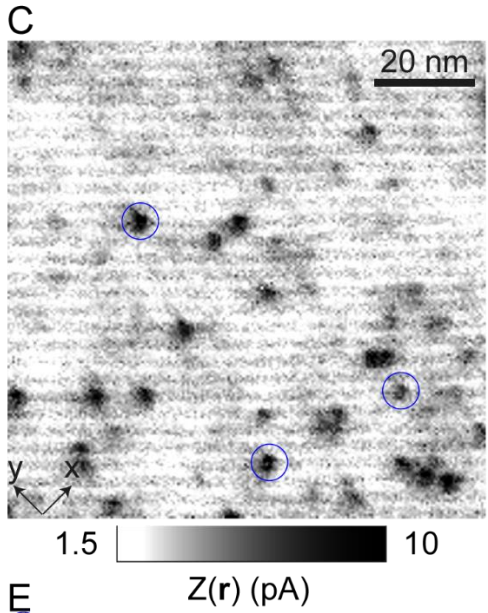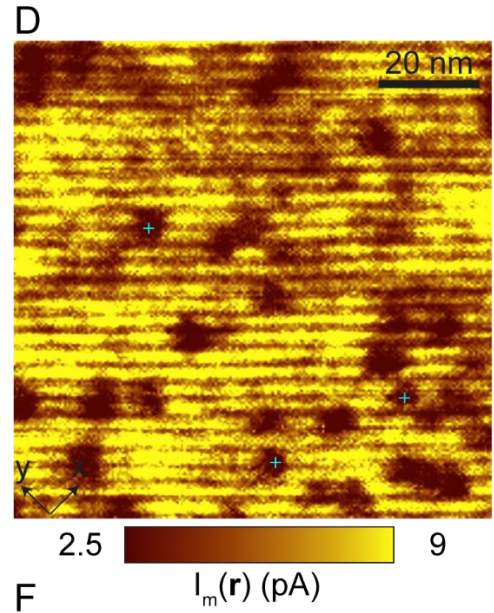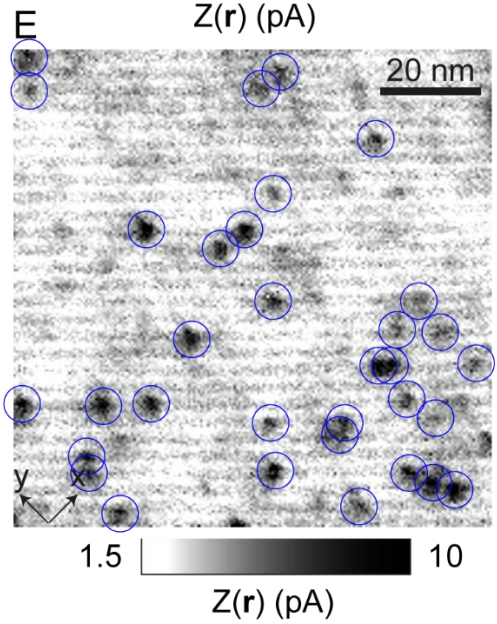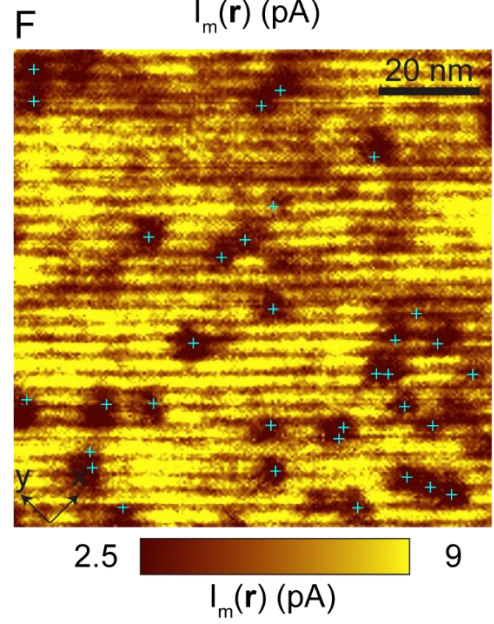

Figure S2. Identification and transformation of Zn impurity sites. (A). Scanning tunnelling spectra on Zn impurity site measured by a superconducting tip with coherence peaks near $E \approx \pm 20 meV$. The resonance peak of Zn impurity moves to energy slightly larger than $\pm 20 meV$. (B). I-V curves measured on Zn impurity site and defect free region. (C). $Z(\boldsymbol{r}) \equiv I(20mV, \boldsymbol{r}) - I(-20mV, \boldsymbol{r})$ map. Three Zn impurities used to generate the transformation matrix are marked by blue circles. (D). Measured $I_m(\mathbf{r})$ map. Three Zn impurities used to generate the transformation matrix are marked by cyan crosses. (E). $Z(\boldsymbol{r}) \equiv I(20mV, \boldsymbol{r}) - I(-20mV, \boldsymbol{r})$ map. Identified Zn impurities are marked by blue circles. (F). Measured $I_m(\mathbf{r})$ map. Red crosses mark the Zn impurities transformed from cyan crosses in (E).

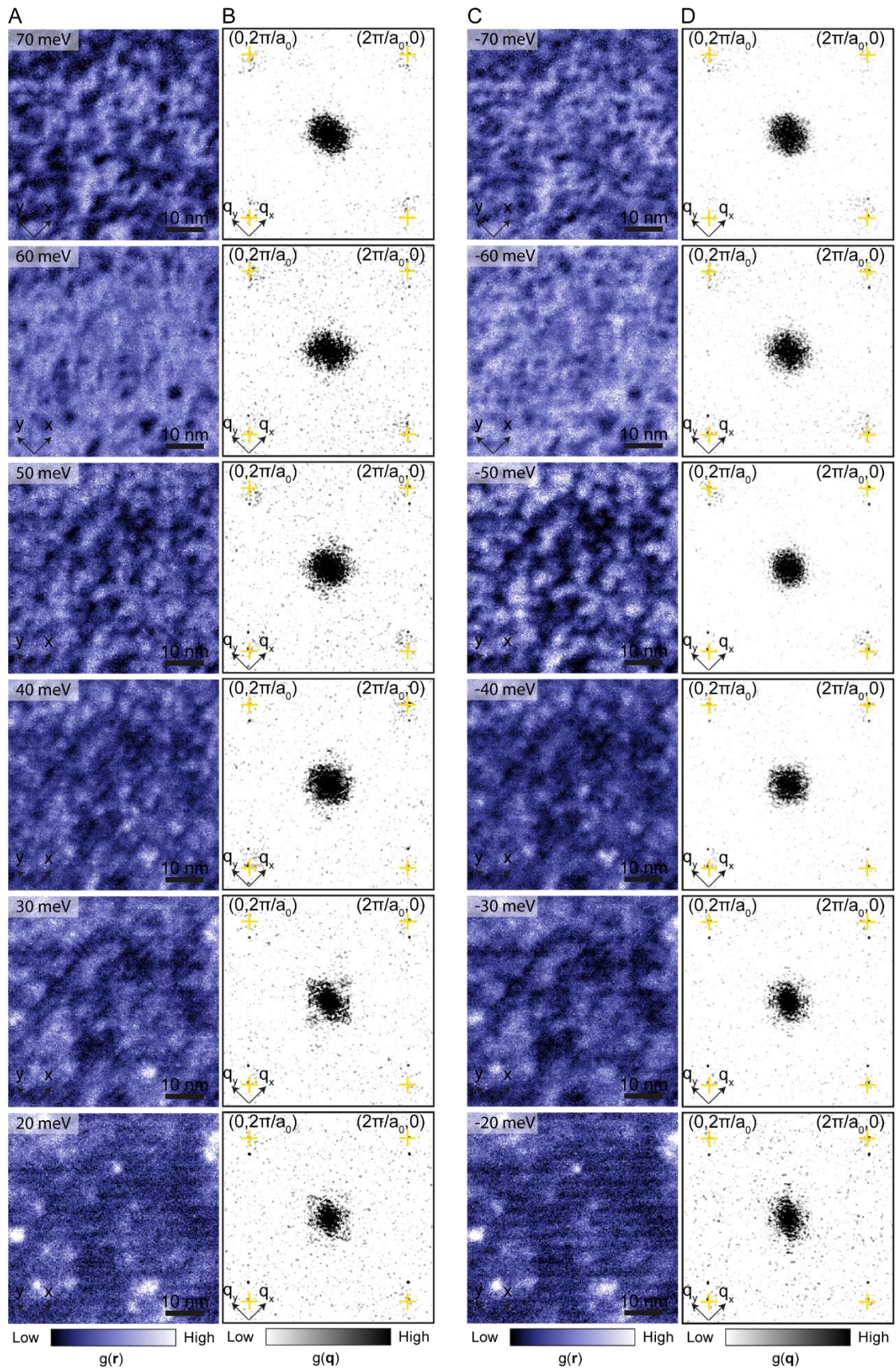

Figure S3. Absence of charge density modulation signature in these samples $p \approx 0.17$ (A,C). Scanning differential tunnelling spectra g(**r**) map at $E = 70\ meV \sim 20 meV$ (A) and $E = -70\ meV \sim -20 meV$ (C). (B,D). Power spectral density Fourier transform of g(**r**), g(**q**) at $E = 70\ meV \sim 20 meV$ (B) and $E = -70\ meV \sim -20 meV$ (D). Bragg peaks $(0, \pm\frac{2\pi}{a_0})$, $(\pm\frac{2\pi}{a_0}, 0)$ are marked by yellow crosses. The pair density wave peaks Q=$(0, \pm\frac{2\pi}{4a_0})$, $(\pm\frac{2\pi}{4a_0}, 0)$ seen in $I_m(\boldsymbol{q})$ map of the main paper cannot be detected throughout g(**q**) map B and D, which implies that there is little or no pre-existent charge density modulation in our sample, consistent with previous studies for this doping range[7] in $Bi_2Sr_2CaCu_2O_8$.